\documentclass[onecolumn,showkeys,preprintnumbers,amsmath,amssymb]{revtex4-2}

\usepackage{graphicx}
\usepackage{dcolumn}
\usepackage{bm}
\usepackage[utf8]{inputenc}
\usepackage{amsmath}
\usepackage{amssymb}
\usepackage{subfigure}
\usepackage{booktabs}
\usepackage{color}
\usepackage{units}	
\usepackage{subcaption}
\usepackage{float}

\usepackage{subfigmat}
\usepackage{changepage}

\begin{document}


\title{Holographic Ricci dark energy in Nonconservative Unimodular Gravity}
\author{Marcelo H. Alvarenga}
 \altaffiliation{}
 \email{marcelo.alvarenga@edu.ufes.br}
\affiliation{Núcleo Cosmo-ufes \& Departamento de Física, UFES, Vitória, ES, Brazil
}%
\author{Júlio C. Fabris}
 \altaffiliation{}
 \email{julio.fabris@cosmo-ufes.org}
\affiliation{Núcleo Cosmo-ufes \& Departamento de Física, UFES, Vitória, ES, Brazil
\\
National Research Nuclear University MEPhI, Kashirskoe sh. 31, Moscow 115409, Russia
}%

\author{Hermano Velten}%
\email{hermano.velten@ufop.edu.br}
\affiliation{Departamento de F\'isica, Universidade Federal de Ouro Preto (UFOP), Campus Universit\'ario Morro do Cruzeiro, 35.400-000, Ouro Preto, Brazil}

\author{Luiz Filipe Guimarães}
\email{luiz.f.guimaraes@ufes.br}
\affiliation{Núcleo Cosmo-ufes \& Departamento de Física, UFES, Vitória, ES, Brazil
}

\date{\today}

\begin{abstract}
The structure of unimodular gravity (UG) is invariant to a subclass of diffeomorphism, the transverse diffeomorphism, due to the unimodular condition ($\sqrt{-g}=\epsilon=cte$). Consequently, there is a freedom to define how the conservation laws of the energy-momentum tensor in unimodular gravity in the cosmological context. One of the main characteristics of the complete system of equations that describe cosmological dynamics in UG is that they form an underdetermined system if the usual conservation law of the energy-momentum tensor is not used in your structure, that is, it is necessary to insert extra information into the system to solve the complete set of equations. In this article, we propose the construction of a background cosmological model based on the description of a holographic dark energy component with a cutoff of the order of Ricci scalar in non-conservative UG. Although this choice is indeed a new addition of information to the cosmological system, the complete set of equations remains underdetermined, however, the new feature of this cosmological model is the appearance of an interaction between matter and dark energy. Indeed, this is a well-known characteristic of cosmological models in which we have holographic dark energy density. Consequently, we propose an ansatz to the interaction term $Q=\beta H \rho_{m}$, and obtain the cosmological parameters of our model. We found a viable universe model with similar characteristics to the $\Lambda \mathrm{CDM}$ model. We performed statistical analysis of the background model using the "Cosmic Chronometer" (CC) data for $H(z)$, and obtain as a result using Akaike Information Criterion (AIC), and the Bayesian Information Criterion (BIC) as model selection criteria that $\Lambda \mathrm{CDM}$ prevails as the best model. However, the proposed model is competitive when compared to the cosmological model $\omega\mathrm{CDM}$.

\end{abstract}

\keywords{Gravitation, Unimodular Gravity, Dark Energy, Holographic Principle.}
\maketitle

\section{Introduction}

At the beginning of the 20th century, a new conception was born in the description of gravitational phenomena that triggered a great advance in the understanding of the laws that govern nature, the theory of General Relativity (GR), published in 1916 by Albert Einstein \cite{Einstein:1916vd}. It describes gravity according to the geometry of space-time in four dimensions. The mathematical structure of GR is based on differential geometry and allows the description of gravitational phenomena on solar system scales as well as the description of the universe as a whole. Regarding the symmetries present in GR, it is invariant under general diffeomorphism, which directly leads to the conservation of the energy-momentum tensor. Over the years and with technological advancements, it was possible to carry out several experiments to test GR predictions. It has passed all the tests it was subjected to, e.g., the detection of the deflection of light that passes close to a gravitational field source, and after a century of the publication of GR, the gravitational waves and black holes, predicted by the theory, were detected \cite{Dyson:1920cwa,LIGOScientific:2016aoc}. 

Within the cosmological context, it is from GR, equipped with a cosmological constant, that we have the most successful description of the universe, the standard cosmological model, $\Lambda\mathrm{CDM}$ \cite{Peebles:2002gy,Zeldovich:1968ehl}. The model explains the accelerated expansion of the universe and it is the best fit to the observational data, even though in its description it introduces a dark sector given with two exotic components of unknown origin at the moment, dark matter and dark energy. The latter is responsible for the accelerated expansion of the universe driven by the cosmological constant $\Lambda$ \cite{Zeldovich:1968ehl,Copeland:2006wr}. However, its success notwithstanding, recent studies point to the existence of limitations in the description of the standard cosmological model \cite{Bull:2015stt}. 
For example, there is a huge discrepancy between the observed value for the cosmological constant, and the theoretical value described by quantum field theory for the vacuum density, which can be estimated up to 120 orders of magnitude difference \cite{Martin:2012bt}. This leads to a fundamental problem in theoretical physics, the problem of the cosmological constant, for details see \cite{Bull:2015stt,Martin:2012bt, RevModPhys.61.1,Copeland:2006wr}. Although the $\Lambda\mathrm{CDM}$ model has an important role in cosmological research, its limitations make it clear that further work in the search for better understanding and explanation is necessary. Consequently, these limitations lead researchers to currently explore, for example, alternative theories consisting of modifications to the GR structure.

A little after the publication of GR, alternative theories began to appear. In 1919, an unimodular version of GR was proposed by Einstein himself and was later known as Unimodular Gravity (UG) \cite{Einstein:1919gv,Anderson:1971pn}. In UG, a constraint is imposed on the determinant of the metric tensor, fixing it to be a constant. One of the striking features of UG is that its field equations are trace-free, and this means that the conservation of the energy-momentum tensor is not a direct consequence of Biachi's identities, as can be seen in more detail in \cite{Ellis:2010uc}. If the conservation of the energy-momentum tensor is considered, we have as a result GR and the addition of a cosmological constant, again referred to as $\Lambda$, which this time is an integration constant. Regarding the symmetries present in UG, they are invariant to a subclass of diffeomorphisms that preserves the determinant of the metric tensor, also known as transverse diffeomorphism. This allows us to review the general conservation laws of the energy-momentum tensor, which has attracted a lot of attention to UG theory, see \cite{Garcia-Aspeitia:2019yod,Fabris_2022,Fabris:2023vop} and references therein for a discussion in the cosmological context. The non-conservative version of UG, which we call Nonconservative Unimodular Gravity ($\mathrm{NUG}$), has interesting implications for cosmology. One of its main consequences is the fact that the equations that describe the background cosmological dynamics form an underdetermined system, which occurs due to the structure of the unimodular theory in which traceless field equations lead to the absence of geometric information such as, for example, the Ricci scalar. Consequently, there is a need to introduce extra information to solve the set of equations in the cosmological description, as studied in \cite{Garcia-Aspeitia:2019yod,Fabris_2022}. 

Following this same line of thought, we propose in this article the construction of a background cosmological model in NUG considering a holographic dark energy component, which we will be addressed here as holographic dark energy in nonconservative unimodular gravity ($\mathrm{HODENUG}$). Holographic models suggest a dynamic approach to dark energy and in this sense, the description from GR leads to a modification to the $\Lambda\mathrm{CDM}$ model.
%
These models are characterized by an existing relationship between the infrared and ultraviolet cutoff in the energy of a physical system, as proposed in Ref.\cite{Cohen:1998zx}, and which will be maintained in the present work. This relationship will ensure that the energy in a given volume does not exceed the energy of a black hole of the same size \cite{Cohen:1998zx,Zimdahl:2007zz}. When dealing with the dynamics of the Universe, the infrared cutoff has to have the same cosmological length scale. Studies proposing different scales of infrared cutoff in the cosmological context of GR can be found in \cite{Zimdahl:2007zz,Li:2004rb,delCampo:2011jp,wang2017holographic}. In the present work, we propose an infrared cutoff proportional to the Ricci scalar for the dark energy density in NUG. In a GR based cosmological scenario, it was proposed for the first time in \cite{PhysRevD.79.043511} and later observational constraints were proposed, for example, in Refs.\cite{Fu:2011ab,delCampo:2013hka}. Firstly, we impose the holographic dark energy density proportional to the Ricci scale as an ansatz in the set of cosmological equations in NUG. However, we find that this mere consideration leads to a cosmological model with interaction between matter and holographic dark energy, in the same way as has been obtained in the context of GR \cite{Benetti_2019}. This means that the underdetermined system of equations still persists in the cosmological description in $\mathrm{HODENUG}$, and consequently, we insert a new condition for this model: extra information under the interaction term. The cosmological models with interaction between matter and dark energy is an alternative proposal due to the problem of the cosmological constant. This interpretation, as pointed out in Ref.\cite{Zimdahl:2012mj} gives rise to richer cosmological dynamics compared to models without interaction. We will compare the background cosmological dynamics in $\mathrm{HODENUG}$ with data from the history $H\left(z\right)$ of the Hubble parameter.

The structure of the paper is as follows. In section (\ref{22}) we will make a general review of the ideas present in the construction of NUG, taking as reference the work described in Ref.\cite{Fabris_2022} and in section (\ref{sec3}) we will give the general ideas present in the holographic description for dark energy, taking as a reference the works in Refs.\cite{Cohen:1998zx,Zimdahl:2007zz}. The main results of this article are presented in Section (\ref{sec4}). In section (\ref{sec4}) we will build the background cosmological model in $\mathrm{HODENUG}$ and find some cosmological parameters for this model. In section (\ref{sec5}) there is the statistical analysis of the comparison with cosmological data for the $\mathrm{HODENUG}$ model, the $\Lambda\mathrm{CDM}$ model and $\omega\mathrm{CDM}$. And finally, in the section (\ref{sec6}) we present final remarks and conclusions.

\section{Overview nonconservative unimodular gravity.}\label{22}

 The unimodular theory is characterized by imposing that the determinant of the metric is a constant, see \cite{Einstein:1919gv, Anderson:1971pn, RevModPhys.61.1, 1991JMP....32.1337N, 1993JMP....34.2465T}. The equations in UG can also be obtained through the Einstein-Hilbert action with the addition of the unimodular condition as a constraint in the system. For details see \cite{Anderson:1971pn, Fabris_2022}. In brief summary, the total action (consisting of the Einstein-Hilbert action plus the action for the matter fields) on UG is
\begin{align}
S_{UG} & =\frac{1}{2\kappa}\int d^{4}x\left[\sqrt{-g}R-\chi\left(\sqrt{-g}-\epsilon\right)\right]+\int d^{4}x\sqrt{-g}\mathcal{L}_{m},\label{eqII:1}
\end{align}
in which $\chi$ can be easily identified as a Lagrange multiplier and $\epsilon$ is a tensor density. 
By varying the above action with respect to the Lagrange multiplier $\chi$, one obtains
\begin{align}
\sqrt{-g} & =\epsilon\label{eqII:2},
\end{align}
which in fact establishes the unimodular constraint. 
On the other hand, varying the total action with respect to the metric tensor $g^{\mu\nu}$, one obtains the following expression
\begin{align}
R_{\mu\nu}-\frac{1}{2}g_{\mu\nu}R +\frac{1}{2}\chi g_{\mu\nu} & =8\pi G T_{\mu\nu}.\label{eqII:3}
\end{align}
Finding the trace of (\ref{eqII:3}), one obtains
\begin{align}
\chi & =\frac{R}{2}+4\pi G T,\label{eqII:4}
\end{align}
Inserting the expression (\ref{eqII:4}) into (\ref{eqII:3}), 
\begin{align}
R_{\mu\nu}-\frac{1}{4}g_{\mu\nu}R & =8\pi G\left(T_{\mu\nu}-\frac{1}{4}g_{\mu\nu}T\right),\label{eqII:5}
\end{align}
which are the field equations trace-free in UG. Using the Bianchi identities in (\ref{eqII:5}) one obtains following expression
\begin{align}
\frac{R^{;\nu}}{4} & =8\pi G\left(T^{\mu\nu}\,_{;\mu}-\frac{1}{4}T^{;\nu}\right).\label{eqII:6}
\end{align}

A striking characteristic of the traceless field equations and, consequently, of the field equations in UG: the Bianchi identities do not directly lead to the conservation of the energy-momentum tensor. This is widely discussed in the literature, e.g. \cite{Ellis:2010uc, Ellis:2013uxa,Fabris_2022,Fabris:2023vop, universe7020038}, and allows to explore the non-conservation of the energy-momentum tensor within the UG structure \cite{Fabris_2022,Garcia-Aspeitia:2019yod}. This leads UG to distinguish itself from GR, especially in the cosmological scenario. However, it should be pointed out that if the conservation of the energy-momentum tensor is imposed on the UG structure we recover GR+$\Lambda$ with $\Lambda$ an integration constant, as previously mentioned. The expression 'non-conservation' may sound misleading, since it does not mean to give up the conservation laws in the strict sense, but as a generalization of the usual, minimal, conservation laws by introducing some new interactions between the matter field and the geometry. It must be kept in mind this remark when using the expression “non-conservation of the energy-momentum tensor”.

In this article, we keep the ideas introduced in \cite{Fabris_2022}, not imposing the conservation of the energy-momentum tensor in the unimodular setting. The structure of equations (\ref{eqII:5}) and (\ref{eqII:6}) are preserved if the holographic dark energy component is introduced. This is done in section (\ref{sec4}). 

In the following subsection, we will give a brief introduction to the background cosmological model in NUG.

\subsection{Brief review of the background cosmological structure in NUG.}\label{sec2}

For a review of the cosmological analysis in NUG, we will consider the homogeneous, isotropic, and expanding cosmological background considering the flat Friedmann–Lemaitre –Robertson–Walker (FLRW) metric given by
\begin{align}
ds^{2} & =dt^{2}-a^{2}\left(t\right)\delta_{ij}dx^{i}dx^{j},\label{eqIIA:1}
\end{align}
where $a\left(t\right)$ is the scalar factor. The cosmological dynamic in NUG described by the metric (\ref{eqIIA:1}) is
\begin{align}
\dot{H} & =-4\pi G\left(\rho+p\right)\label{eqIIA:2}\\
\ddot{H}+4H\dot{H} & =-4\pi G\left(\dot{\rho}+\dot{p}+4H\left(\rho+p\right)\right),\label{eqIIA:3}
\end{align}
where $\rho$ and $p$ are the energy density and pressure, respectively;
$H=\dot{a}/a$ is Hubble parameter. 

With a direct inspection of equations (\ref{eqIIA:2}) and (\ref{eqIIA:3}) one realise that this set of equations is incomplete, i.e, we have no information regarding the first Friedmann equation, related to the Hubble parameter $H$, the constraint equation. They form an underdetermined system of equations, i.e. to solve them it is necessary to introduce an ansatz. A viable model is obtained with the introduction of the ansatz $\bar{\rho}\equiv(\rho+p)\propto a^{-n}$. An example is by fixing $n=4$ leading to the typical behavior of a radiation fluid as discussed in Ref. \cite{Fabris_2022}.

Basically, the introduction of the ansatz described above, with $n = 4$, leads to a split of expression (\ref{eqIIA:3}) into two new equations,

\begin{align}
\frac{d}{dt}\left(\dot{H}+2H^{2}\right) & =0,\label{eqIIA:4}\\
\dot{\bar{\rho}}+4H\bar{\rho} & =0.\label{eqIIA:5}
\end{align}

The solution of equation (\ref{eqIIA:5}) is just the introduced ansatz for the matter component. On the other hand, the solution of equation (\ref{eqIIA:4}) is

\begin{align}
\dot{H}+2H^{2} & =\frac{2}{3}\Lambda_{U}.\label{eqIIA:6}
\end{align}
where $\Lambda_{U}$ is a constant of integration that is associated with the cosmological constant \cite{Fabris_2022} and the term $2/3$ was introduced for convenience. Introducing expression (\ref{eqIIA:6}) into expression (\ref{eqIIA:3}) we obtain

\begin{align}
H^{2} & =\frac{8\pi G}{3}\left[\frac{3}{4}\bar{\rho}+\frac{\Lambda_{U}}{8\pi G}\right],\label{eqIIA:7}
\end{align}
similar to the first Friedmann equation. We must emphasize that for any fluid considered as a constituent of the universe, $\bar{\rho}$ will always have the typical behavior of a radiation fluid \cite{Fabris_2022}. 
Rewriting the expression (\ref{eqIIA:7}) in terms of the fractionary energy density parameter and using the fact of the typical behavior of a radiation fluid to $\bar{\rho}$, we obtain

\begin{align}
H\left(a\right) & =H_{0}\left(\Omega_{U}+\frac{1-\Omega_{U}}{a^{4}}\right)^{\frac{1}{2}},\label{eqIIA:8}
\end{align}
where $\Omega_{U}\equiv\frac{\Lambda_{U}}{3H_{0}^{2}}$ and we use $1=\frac{3}{4}\bar{\Omega}+\Omega_{U}$. We should point out that the expression for the Hubble parameter has a similar structure to the flat $\Lambda\mathrm{CDM}$ model without pressureless components. However, the term $(1-\Omega_{U})=3\bar{\Omega}/4$ has a different physical interpretation from the cosmological constant in $\Lambda\mathrm{CDM}$. All types of fluid, whether relativistic or non-relativistic, must be added to this expression.
In \cite{Fabris_2022}, for $\Lambda_{U}>0$, the deceleration parameter $q$ and age of the universe today $t_{0}$ were calculated and the estimates led to a description of the universe that transitions from a radiative phase evolving to an accelerated de Sitter expansion. The age of the universe and the estimated value for the cosmological constant $\Lambda_{U}$ were also obtained and present values similar to the $\Lambda\mathrm{CDM}$ model \cite{Fabris_2022}.


   
Basically, using Hubble parameter, expression (\ref{eqIIA:8}), the deceleration parameter for this model is

\begin{align}
    q\left(a\right)	=\frac{2\left(1-\Omega_{U}\right)a^{-4}}{\left[\Omega_{U}+\left(1-\Omega_{U}\right)a^{-4}\right]}-1.\nonumber
\end{align}


In terms of the redshift $z$, 

\begin{align}
    q\left(z\right)	=\frac{2\left(1-\Omega_{U}\right)\left(1+z\right)^{4}}{\left[\Omega_{U}+\left(1-\Omega_{U}\right)\left(1+z\right)^{4}\right]}-1.\label{eqIIA:9}
\end{align}

At redshift $z=0$, the deceleration parameter today $q_{0}$ is
 
\begin{align*}
    q_{0}	=2\left(1-\Omega_{U}\right)-1.
\end{align*}

The condition for an accelerated universe $q_{0}<0$ at $z=0$ is
\begin{align}
q_{0}<0 & \longrightarrow \Omega_{U} >\frac{1}{2}.\label{eqIIA:10}
\end{align}




We have seen that the background cosmological description in NUG is strictly conditioned on the addition of extra information to solve the system of equations. According to the ansatz introduced in \cite{Fabris_2022} a model of the universe that evolves directly from a direct radiative phase to a de Sitter expansion was found. 


Following this line of reasoning, we propose to maintain the form of the equations (\ref{eqIIA:2}) and (\ref{eqIIA:3}), in the construction of a universe model based on the description
of a component Holographic Dark Energy in Nonconservative Unimodular Gravity ($\mathrm{HODENUG}$).


\section{Overview in holographic dark energy.}\label{sec3}

The holographic dark energy model is based on the holographic principle,
which relates the number of degrees of freedom of a bounded system with
the area of the boundary \cite{Susskind:1994vu,tHooft:1993dmi}. To describe a model for the universe in which the dark energy density is holographic we will build
on the ideas set out in Ref. \cite{Cohen:1998zx}. Basically,
in \cite{Cohen:1998zx} it is established a relation (limit) between the energy
in a box length L with the energy of a black hole, given by the following
relation

\begin{align}
L^{3}\Lambda^{4} & \leq LM_{Pl}^{2},\label{eq:11S}
\end{align}
where $M_{Pl}=(8\pi G)^{-1/2}$ is the reduced Planck mass
and $\Lambda$ is associated with the vacuum energy (ultraviolet cut-off).
The above inequality must be satisfied and excludes all states where
the Schwarzschild radius exceeds the length $L$. Therefore, when considering
box length $L$ as infrared cut-off and $\rho_{H}$ as ultraviolet
cut-off and related to $\Lambda^{4}$, one obtains \cite{Cohen:1998zx,Zimdahl:2007zz}

\begin{align}
L^{3}\rho_{H} & \leq LM_{Pl}^{2}.
\end{align}

Thus, the
largest $L$ allowed will be the one that saturates the inequality
above, so

\begin{align}
\rho_{H} & =\frac{3c^{2}M_{Pl}^{2}}{L{{}^2}},\label{eq:3}
\end{align}
where $c^{2}$ is a dimensionless numerical constant  and the factor $3$ was introduced for convenience. The interest in studying the dynamics of the Universe means that the length $L^2$ has the same cosmological length dimensions, and therefore can be attributed in the form of ansatz.  A detailed study of various models of holographic dark energy (same as different ansatz for $L^2$) in the context of GR can
be found at \cite{Li:2004rb,delCampo:2011jp,wang2017holographic}.

A direct inspection of the Ricci scalar shows that we have that $\dim\,R=L^{-2}$  and consequently the same dimensions as the holographic dark energy density $\rho_H$ (\ref{eq:3}). Holographic Ricci dark energy models were first considered by \cite{PhysRevD.79.043511} applying them to GR based cosmological
dynamics. In this reference the holographic dark energy density $\rho_{H}$ is of the order of the Ricci scalar in NUG.

\section{Holographic Ricci dark energy in NUG}\label{sec4}

The holographic Ricci dark energy density is written as \cite{PhysRevD.79.043511}
 
\begin{align}
\rho_{H} & =-\frac{3c^{2}M_{Pl}^{2}}{6}R=\frac{3c^{2}}{8\pi G}\left(2H^{2}+\dot{H}\right),\label{eq:4}
\end{align}
being the Ricci scalar $R=-6\left(2H^{2}+\dot{H}\right)$.
Differentiating the expression (\ref{eq:4}) with respect to $t$,

\begin{align}
\dot{\rho}_{H} & =\frac{3c^{2}}{8\pi G}\left(4H\dot{H}+\ddot{H}\right).\label{eq:5}
\end{align}

Using the expressions (\ref{eq:4}) and (\ref{eq:5}) in cosmological
dynamics described by NUG, expressions (\ref{eqIIA:2}) and (\ref{eqIIA:3}), one obtains

\begin{align}
3H^{2} & =8\pi G\left(\frac{3}{4}\left(\rho+p\right)+\frac{1}{2c^{2}}\rho_{H}\right),\label{eq:5.2}\\
\dot{\rho}_{H} & =-\frac{3c^{2}}{2}\left(\dot{\rho}+\dot{p}+4H\left(\rho+p\right)\right).\label{eq:5.3}
\end{align}

The expression (\ref{eq:5.2}) is the modified Friedmann equation for
our model, while (\ref{eq:5.3}) is the equation arising from the non-conservation of the energy-momentum tensor.

We consider the combination $\left(\rho+p\right)$ consisting of the non-relativistic matter, i.e.
\begin{align}
\left(\rho+p\right) =\rho_{m}+p_{m}=\rho_{m},
\end{align}
where subscript $m$ refers to non-relativistic matter and we use the pressureless equation of state parameter $p_{m}=\omega_{m}\rho_{m}=0$. With this consideration, the modified Friedmann equation, expression (\ref{eq:5.2}) and non-conservation equation, expression (\ref{eq:5.3}) read
\begin{align}
3H^{2} & =8\pi G\left(\frac{3}{4}\rho_{m}+\frac{1}{2c^{2}}\rho_{H}\right).\label{eq:10.1}\\
\frac{2}{3c^{2}}\dot{\rho}_{H} & =-\left[\dot{\rho}_{m}+4H\rho_{m}\right].\label{eq:10.2}
\end{align}

We can rewrite the expression (\ref{eq:10.2}) into two new equations, splitting the contributions from non-relativistic matter and holographic dark energy. Therefore, the complete set of equations describing cosmological dynamics in $\mathrm{HODENUG}$ are

\begin{align}
    3H^{2} &=8\pi G\left(\frac{3}{4}\rho_{m}+\frac{1}{2c^{2}}\rho_{H}\right),\label{eq1:10.1}\\
\frac{2}{3c^{2}}\dot{\rho}_{H} & =-Q,\label{eq1:11}\\
\dot{\rho}_{m}+&4H\rho_{m}  =Q,\label{eq1:12}
\end{align}
in which $|Q|$ is the magnitude of the interaction term between matter and holographic dark energy. Since there is no well motivated description at the microphsyics level for the interaction term $Q$, it must be implemented in a phenomenological way. A review of interacting cosmological models for different interaction terms in the context of GR can be found in Ref.\cite{doi:10.1142/S0218271815300074}.

Now, we assume the conservation of the energy-momentum tensor in the unimodular structure. If the conservation of the energy-momentum tensor is assumed in (\ref{eqII:6}), the equations governing the background cosmological dynamics become

\begin{align}
   3H^{2}&=8\pi G \rho +\Lambda_{U}\label{eq1:13}\\
    \dot{\rho}+&3H\left(\rho+p\right)=0.\label{eq1:14}
 \end{align}
with $\Lambda_{U}$ the same integration constant as previously detailed. The expressions in (\ref{eq1:13}) and (\ref{eq1:14}) are the same equations that describe cosmological dynamics according to GR+$\Lambda$, as expected.
 
In a sense, these equations with $\Lambda_{U}=0$ are identical to those used in the study of models of the universe with an interaction between matter and dark energy, the latter being added externally through the holographic principle, for example, see \cite{Zimdahl:2007zz,PhysRevD.73.123510} for details in the context of GR. In this new condition, expressions (\ref{eq1:13}) and (\ref{eq1:14}) become 

 \begin{align}
     3H^{2}	&=8\pi G\left(\rho_{m}+\rho_{H}\right)\label{eq1:15}\\
\dot{\rho}_{m}	&+3H\rho_{m}=Q\label{eq1:16}\\
\dot{\rho}_{H}	+3H&\left(1+\omega\right)\rho_{H}=-Q\label{eq1:17},
 \end{align}
where $\rho_{H}$ is holographic dark energy density and $\omega=p_H/\rho_H$ is the equation-of-state parameter for holographic dark energy. All analysis and discussion in which we have a holographic dark energy component with a Ricci scale cutoff within the GR structure can be found in Refs. \cite{delCampo:2013hka,Fu:2011ab}. We highlight that, although the equations for the conservative case lead to a similar cosmology to the case of interactions between matter and holographic dark energy added to GR, a precise comparison with the work developed here can not be done directly. This is due to some technical features in the equations of the $\mathrm{HODENUG}$ cosmological model, such as the factor $4H\rho_{m}$ that appears on the left side of (\ref{eq1:12}), instead of $3H \rho_{m}$ as in (\ref{eq1:16}). Such differences are related to the structure of the theories resulting from assuming or not the usual conservation laws. Even though, we can compare the final parameter estimations in each case, as it will be done later.

Returning to the description of the cosmological model in $\mathrm{HODENUG}$, although our choice in holographic dark energy density represents extra information to cosmological dynamics it is not sufficient to solve the set of equations. The non-conservation equation naturally results in an interaction in the dark sector, represented by the quantity $Q$, expressions (\ref{eq1:11}) and (\ref{eq1:12}).  Consequently, we have three independent equations for four unknown functions, they are $H$, $\rho_{m}$, $\rho_{H}$, and $Q$. This means that we still need additional information to solve the set of equations. Let's choose a condition for the interaction term $Q$ in the next section.

\subsection{The interaction $Q=\beta H \rho_{m}$}

In this subsection, we will fix the value of the interaction term in $Q=\beta H \rho_{m}$ by combining it with the expressions that govern the background cosmological dynamics in $\mathrm{HODENUG}$. Replacing the interaction term $Q$ in expressions (\ref{eq1:11}) and (\ref{eq1:12}), we obtain
\begin{align}
    \frac{2}{3c^{2}}\dot{\rho}_{H} & =-\beta H\rho_{m}\label{eqA:1}\\
\dot{\rho}_{m}+&3H\left(\frac{4}{3}-\frac{\beta}{3}\right)\rho_{m}  =0,\label{eqA:2}
\end{align}
where $\beta$ is a free parameter constant dimensionless. The expression (\ref{eqA:2}) can be easily integrated by obtaining
\begin{align}
    \rho_{m}=\rho_{m0}a^{-\left(4-\beta\right)},\label{eqA:3}
\end{align}
where $\rho_{m0}$ represent the energy density of matter today. Using the solution (\ref{eqA:3}) in (\ref{eqA:1}), we obtain
\begin{align}
    \rho_{H}=\frac{3c^{2}\,\beta\,\rho_{m0}}{2\left(4-\beta\right)}\left[a^{-\left(4-\beta\right)}-1\right]-A\label{eqA:4},
\end{align}
where we defined $a_{0}=1$ and the term $A$ is another integration constant. Using the condition $\rho_{H}\left(a=a_{0}\right)=\rho_{H0}$, we obtain
$\rho_{H0}=-A,$
where $\rho_{H0}$ represent the holographic dark energy density today.
Replacing into expression (\ref{eqA:4}), we get
\begin{align}
    \rho_{H}=\rho_{H0}+\frac{3c^{2}\,\beta\,\rho_{m0}}{2\left(4-\beta\right)}\left[a^{-\left(4-\beta\right)}-1\right].\label{eqA:5}
\end{align}

Note that, in the non-interaction limit $\beta=0$ we obtain the model described in
\cite{Fabris_2022}, where $\rho_{H}=\rho_{H0}=cte$ and $\rho_{m}=\rho_{m0}a^{-4}$. Fixing $\beta=1$, for example, the energy density $\rho_{m}$ behaves like matter (same as described by the $\Lambda\mathrm{CDM}$ model). In the past ($a \rightarrow 0$) the holographic dark energy density behaves as non-relativistic matter, while it evolves to a cosmological constant for a distant future ($a \rightarrow \infty$). In the next subsection, we write the cosmological parameters taking into account the behavior of energy densities.

\subsection{Cosmological Parameters}

Using the expressions (\ref{eqA:3}) and (\ref{eqA:5}) we can rewrite
the expression for Hubble's parameter (\ref{eq1:10.1}), in terms of the redshift parameter $z$, as
\begin{align}
    H^{2}\left(z\right)	=H_{0}^{2}\left\{ \left[\frac{3}{4}\Omega_{m0}+\frac{3\,\beta}{4\left(4-\beta\right)}\Omega_{m0}\right]\left(1+z\right)^{\left(4-\beta\right)}+\frac{\Omega_{H0}}{2c^{2}}-\frac{3\,\beta}{4\left(4-\beta\right)}\Omega_{m0}\right\},\label{eqB:1}
\end{align}
where $a=\left(1+z\right)^{-1}$, $H_0$ is the Hubble's parameter today and $\Omega_{i0}=8\pi G \rho_{i0}/3H_{0}^{2}$ is the fractionary energy density parameter today for each component. The dimensionless Hubble parameter becomes
\begin{align}
   E^2 \left(z\right)=\frac{H^{2}\left(z\right)}{H_{0}^{2}}	=\left\{ \left[\frac{3}{4}\Omega_{m0}+\frac{3\,\beta}{4\left(4-\beta\right)}\Omega_{m0}\right]\left(1+z\right)^{\left(4-\beta\right)}+\frac{\Omega_{H0}}{2c^{2}}-\frac{3\,\beta}{4\left(4-\beta\right)}\Omega_{m0}\right\}.\label{eqB:2}
\end{align}

The deceleration parameter $q\left(z\right)$ of our model is given by 
\begin{align}
q\left(z\right)	&=\frac{H'\left(z\right)}{H\left(z\right)}\left(1+z\right)-1\nonumber\\
	&=\frac{\left(4-\beta\right)}{2}\frac{\left[\frac{3}{4}\Omega_{m0}+\frac{3\,\beta}{4\left(4-\beta\right)}\Omega_{m0}\right]\left(1+z\right)^{\left(4-\beta\right)}}{\left[ \left(\frac{3}{4}\Omega_{m0}+\frac{3\,\beta}{4\left(4-\beta\right)}\Omega_{m0}\right)\left(1+z\right)^{\left(4-\beta\right)}+1-\frac{3}{4}\Omega_{m0}-\frac{3\,\beta}{4\left(4-\beta\right)}\Omega_{m0}\right]}-1.\label{eqB:3}
\end{align}
In the above expression, the prime $(')$ represents derivative with respect to the redshift $z$. The deceleration parameter today $q_0$ is

\begin{align}
    q_{0}=	\frac{\left(4-\beta\right)}{2}\left[\frac{3}{4}\Omega_{m0}+\frac{3\,\beta}{4\left(4-\beta\right)}\Omega_{m0}\right]-1,\label{eqB:4}
\end{align}
where we are using the identity $1=\frac{3}{4}\Omega_{m0}+\frac{\Omega_{H_0}}{2c^{2}}$. Current cosmological observations show that our universe experiences an accelerated expansion phase, which leads to a negative decelerating parameter, $q_{0}<0$.
Therefore, this leads to the following condition for the parameter $\Omega_{m0}$

\begin{align}
q_{0}<0 & \longrightarrow \Omega_{m0} <\frac{2}{3}.\label{eqB:5}
\end{align}

In the next section, we will carry out the statistical analysis of this model using cosmological data about the cosmological expansion rate.

\section{Observational constraints using $H\left(z\right)$} \label{sec5}
The Hubble parameter can be written in terms of the redshift $z$ as

\begin{align*}
H\left(z\right) & =-\frac{1}{1+z}\frac{dz}{dt}.
\end{align*}

Therefore, to determine the value of $H\left(z\right)$ we must know
the derivative of z. We can use the age differential technique to
obtain the redshift variation with respect to cosmic time, known as
the Cosmic Chronometer (CC). The technique
consists of measuring the age difference between two galaxies that
formed practically at the same time and that present small variations
in redshift $z$. This technique was proposed by \cite{Jimenez:2001gg}. We will
use the data listed in the table (\ref{tab:CC}).

\begin{figure}[!b]
\centering
\includegraphics[scale=0.4]{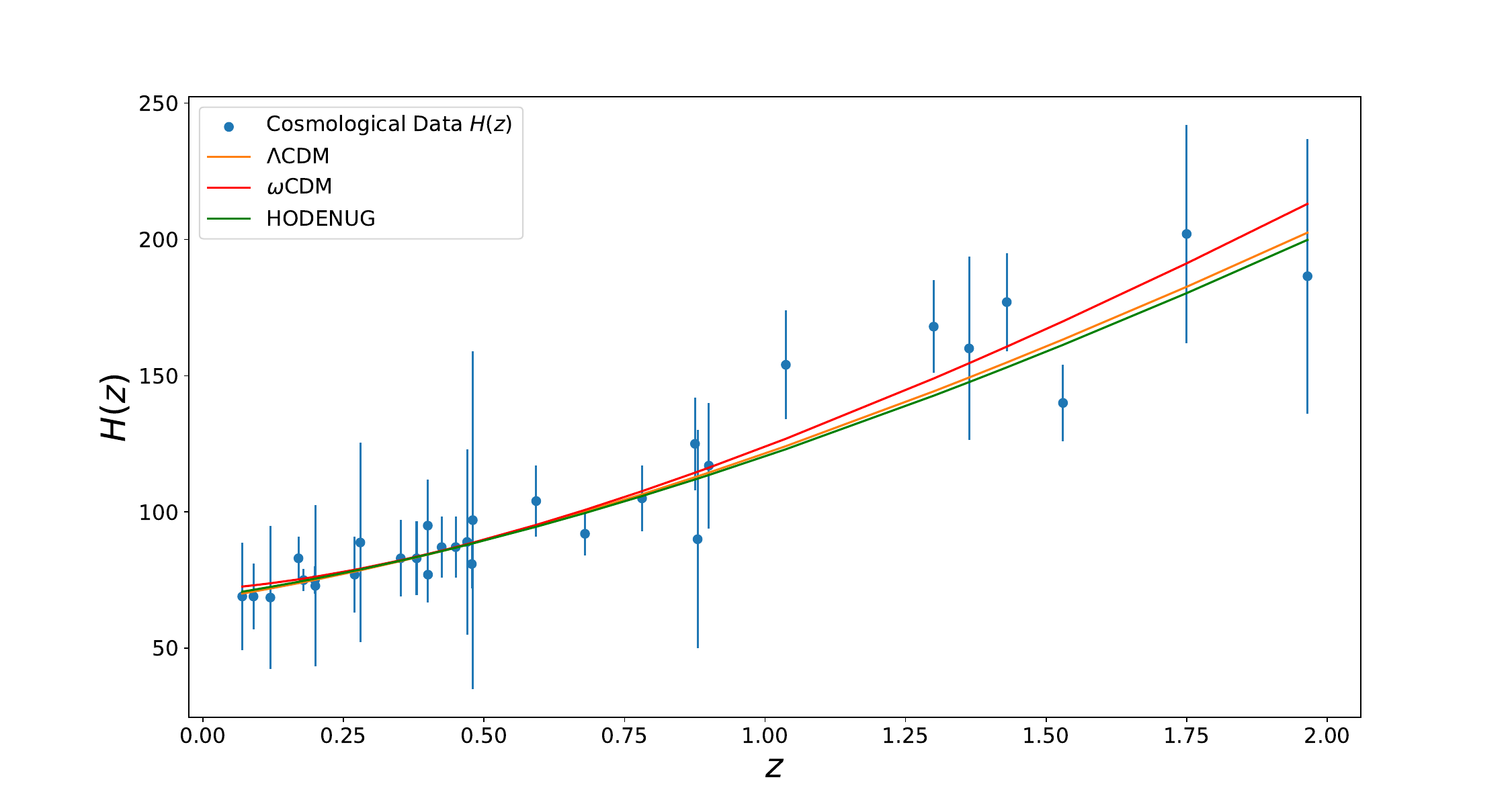}
\caption{Evolution of the Hubble parameter in three cases: the orange graph corresponds to the $\Lambda \mathrm{CDM}$ model, the red graph corresponds to the $\omega\mathrm{CDM}$ model and the green graph corresponds to our model. The blue-colored data are the observed values of H(z).}\label{fig:LambdaCDMHDE}
\end{figure}
\begin{table}[h]
\centering
    \caption{CC data used for analysis.}
    \label{tab:CC}
\resizebox{15cm}{!}{
\begin{tabular}{|cccc|cccc|cccc|cccc|}
\toprule
$z$ & $H\left(z\right)$ & $\sigma_{i}$ & $Ref.$ & $z$ & $H\left(z\right)$ & $\sigma_{i}$ & $Ref.$ & $z$ & $H\left(z\right)$ & $\sigma_{i}$ & $Ref.$ & $z$ & $H\left(z\right)$ & $\sigma_{i}$ & $Ref.$\tabularnewline
\midrule
$0.07$ & $69$ & $19.6$ &  \cite{29} & $0.28$ & $88.8$ & $36.6$ &  \cite{29} & $0.4783$ & $80.9$ & $9$ &  \cite{31} & $1.037$ & $154$ & $20$ & \cite{28} \tabularnewline
$0.09$ & $69$ & $12$ &  \cite{27} & $0.3519$ & $83$ & $14$ &  \cite{28} & $0.48$ & $97$ & $62$ &  \cite{27} & $1.3$ & $168$ & $17$ & \cite{28}\tabularnewline
$0.12$ & $68.6$ & $26.2$ &  \cite{29} & $0.3802$ & $83$ & $13.5$ &  \cite{31} & $0.5929$ & $104$ & $13$ &  \cite{28} & $1.363$ & $160$ & $33.6$ & \cite{30}\tabularnewline
$0.17$ & $83$ & $8$ &  \cite{27} & $0.4$ & $95$ & $17$ & \cite{27} & $0.6797$ & $92$ & $8$ &  \cite{28} & $1.43$ & $177$ & $18$ &  \cite{27}\tabularnewline
$0.1791$ & $75$ & $4$ &  \cite{28} & $0.4004$ & $77$ & $10.2$ &  \cite{31} & $0.7812$ & $105$ & $12$ &  \cite{28} & $1.53$ & $140$ & $14$ &  \cite{27}\tabularnewline
$0.1993$ & $75$ & $5$ &  \cite{28} & $0.4247$ & $87.1$ & $11.2$ & \cite{31} & $0.8754$ & $125$ & $17$ &  \cite{28} & $1.75$ & $202$ & $40$ &  \cite{27}\tabularnewline
$0.2$ & $72.9$ & $29.6$ &  \cite{29} & $0.4497$ & $87.1$ & $11.2$ &  \cite{31} & $0.88$ & $90$ & $40$ &  \cite{27} & $1.965$ & $186.5$ & $50.4$ & \cite{30}\tabularnewline
$0.27$ & $77$ & $14$ &  \cite{27} & $0.47$ & $89$ & $34$ &  \cite{32} & $0.9$ & $117$ & $23$ &  \cite{27} &  &  &  & \tabularnewline
\bottomrule
\end{tabular}
}
\end{table}

Using the usual $\chi^{2}$ statistic, we calculate the following quantity
\begin{align}
\chi^{2} & =\sum_{i}\frac{\left(H^{t}\left(z_{i}\right)-H^{o}\left(z_{i}\right)\right)^{2}}{\sigma_{i}^{2}},\label{eqo:1}
\end{align}
where the subscript $t$ and $o$ corresponds to the theoretical and observational value of $H\left(z_{i}\right)$, while $\sigma_{i}$ correspond to the uncertainty attributed to each value of $H^{o}\left(z_{i}\right)$.
From this quantity, we can construct the probability distribution function (PDF)

\begin{align}
\mathcal{P} & =\mathcal{A}\exp(-\frac{\chi^{2}}{2}),\label{eqo:1.1}
\end{align}
where $\mathcal{A}$ it is normalization factor. The expression (\ref{eqo:1.1}) depends on the free parameters of the theoretical model. For our $\mathrm{HODENUG}$ model, we can rewrite the Hubble parameter, expression (\ref{eqB:1}), in terms of three free parameters, $\Omega_{m0}$, $\beta$ and $h$. Therefore, the Hubble parameter of the theoretical model is

\begin{align}
     H\left(z,\Omega_{m0},\beta,h\right)	=100\,h\left[ \left(\frac{3}{4}\Omega_{m0}+\frac{3\,\beta}{4\left(4-\beta\right)}\Omega_{m0}\right)\left(1+z\right)^{\left(4-\beta\right)}+1-\frac{3}{4}\Omega_{m0}-\frac{3\,\beta}{4\left(4-\beta\right)}\Omega_{m0}\right]^{\frac{1}{2}}.\label{eqo:1}
\end{align}

It is important, in statistical analysis, to choose a prior appropriately.
This leads us to correctly choose the intervals in which the free
parameters are established. In our model, we must have

\begin{align*}
0\leq h\leq1;\,\,0.01\leq\Omega_{m0}\leq0.75\,\, & -0.85\leq\beta\leq3.50.
\end{align*}

\begin{figure}[h!]
    \centering
   \subfigure[]{\includegraphics[scale=0.35]{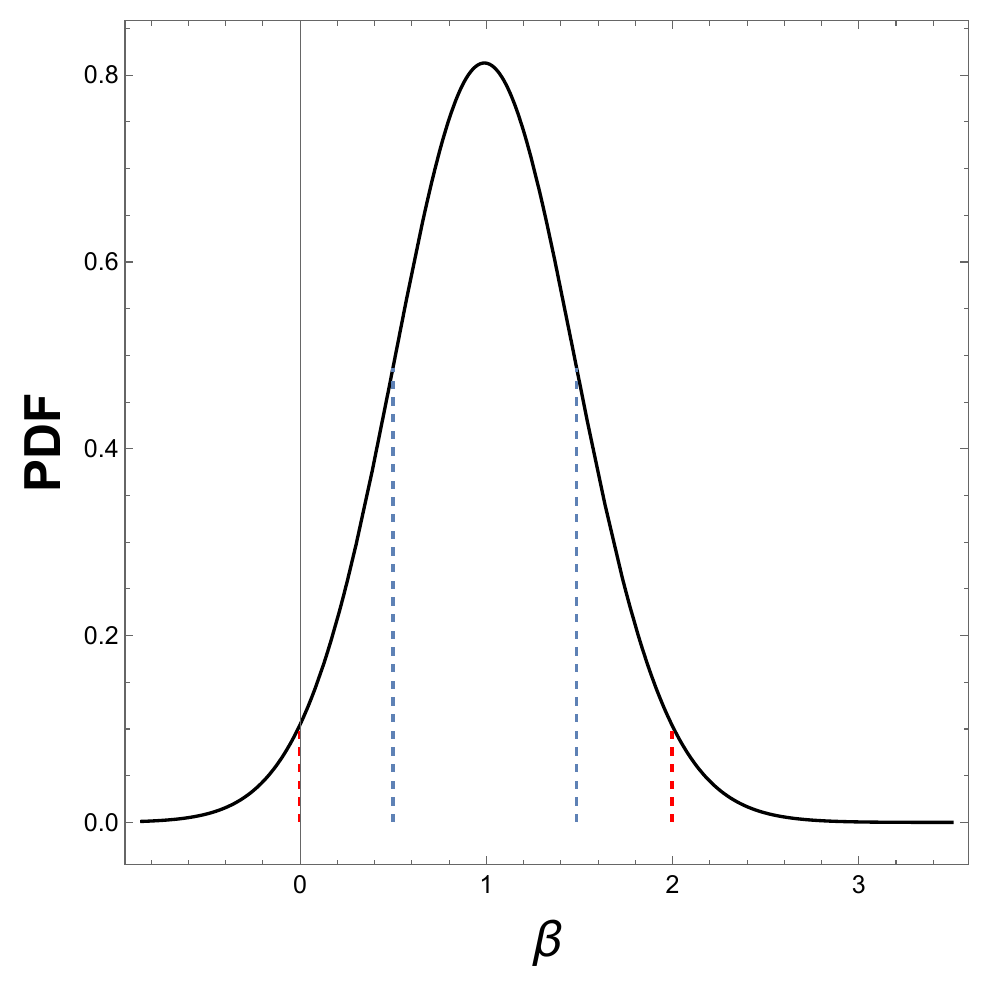}\label{fig:Graficos5}}\ \ \ \ \ \ \
   \subfigure[]{\includegraphics[scale=0.35]{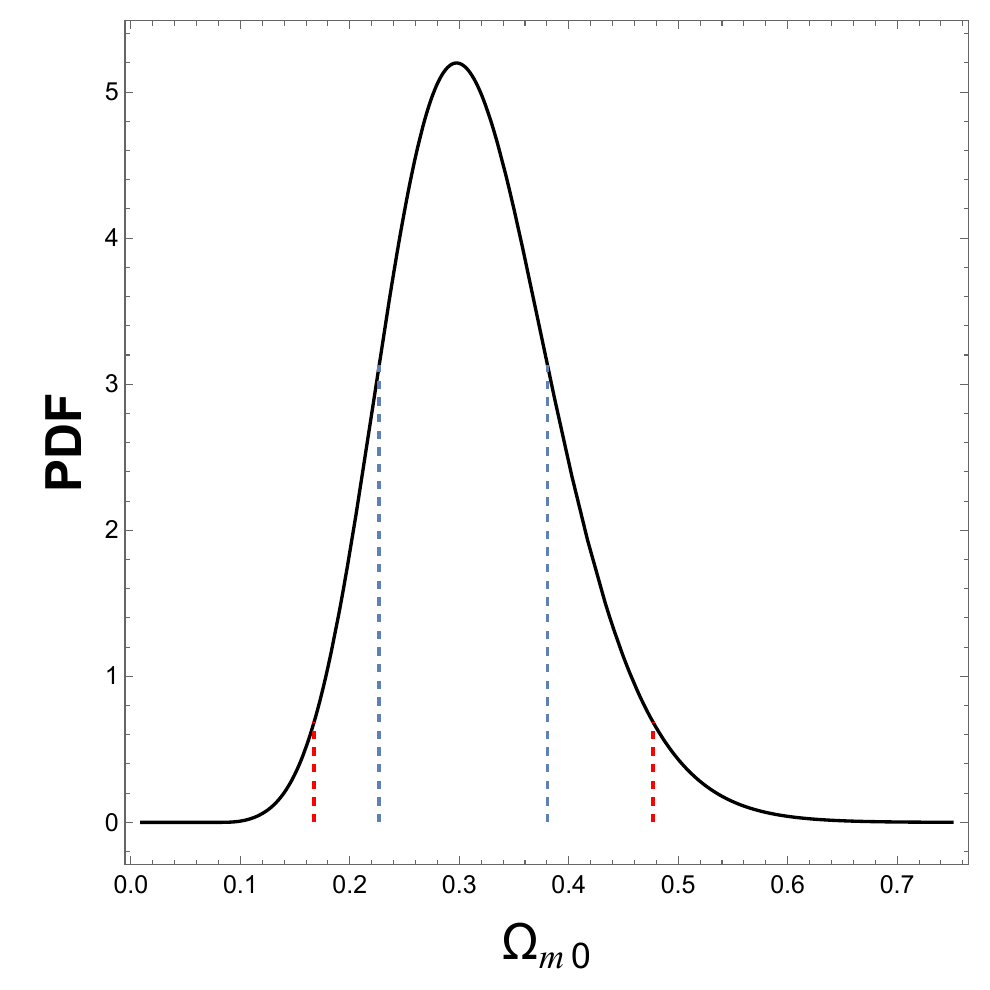}\label{fig:Graficos6}}
\caption{The one-dimmensional PDFs for $\beta$ and $\Omega_{m0}$. Dashed lines in \subref{fig:Graficos5} and \subref{fig:Graficos6} represent $68.2\%$ (blue color) and $95.4\%$ (red color) confidence level, respectively.}
\label{fig:baseline}
   \end{figure}

In our subsequent analysis we fixed a value for $h$ so that we obtain the minimum value of $\chi^{2}$, that is, $h=0.685$. The one-dimensional PDF of
the $\beta$ and $\Omega_{m0}$ parameters can be found in the Fig.(\ref{fig:baseline}). The estimations we have obtained are: $\Omega_{m0}=0.297^{+0.083}_{-0.071}$ and $\beta=0.989^{+0.495}_{-0.491}$ at the $68.2\%$ confidence level, $\Omega_{m0}=0.297^{+0.179}_{-0.130}$ and $\beta=0.989^{+1.008}_{-0.991}$ at the $95.4\%$ confidence level as shown in Fig.(\ref{fig:baseline}).

In this paper, we use two background cosmological models for comparison: the $\Lambda\mathrm{CDM}$ and $\omega\mathrm{CDM}$ models. The Hubble parameter $H$ used for the flat $\Lambda\mathrm{CDM}$ model is described as
\begin{align}
    H\left(z,\Omega_{m0},h\right)	=100h\left[\Omega_{m0}\left(1+z\right)^{3}+\left(1-\Omega_{m0}\right)\right]^{\frac{1}{2}}.
\end{align}
On the other hand, the Hubble parameter used for the flat $\omega\mathrm{CDM}$ model is described as
\begin{align}
    H\left(z,\Omega_{m0},h,\omega\right)	=100h\left[\Omega_{m0}\left(1+z\right)^{3}+\left(1-\Omega_{m0}\right)\left(1+z\right)^{3\left(1+\omega\right)}\right]^{\frac{1}{2}}.
\end{align}
The evolution of the Hubble parameter in comparison with the $\Lambda \mathrm{CDM}$ and $\omega\mathrm{CDM}$ model is shown in Fig.(\ref{fig:LambdaCDMHDE}). In the plot, we have used the best fit values for the free parameters for each model as shown in the table (\ref{tab:LCDMHDENUG}).

\begin{figure}[t!]
\centering
   \subfigure[]{\includegraphics[scale=0.6]{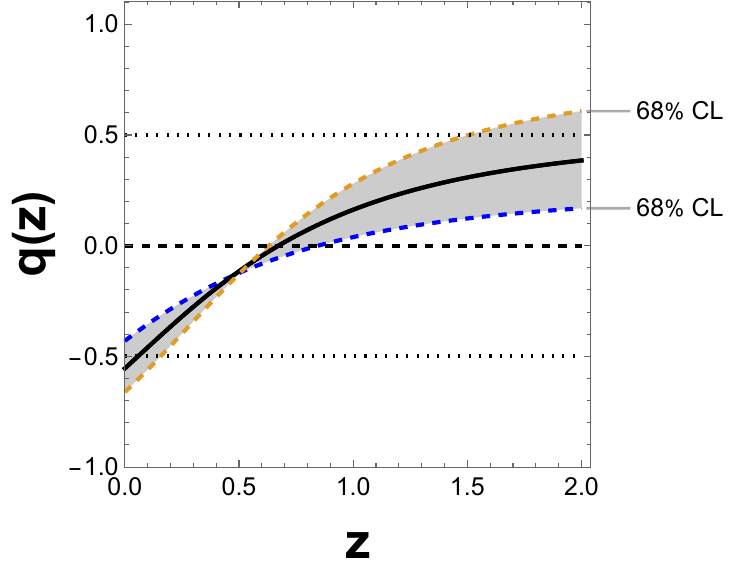} \label{fig:qz68}} \ \ \ \ \ \ \
\subfigure[]{\includegraphics[scale=0.6]{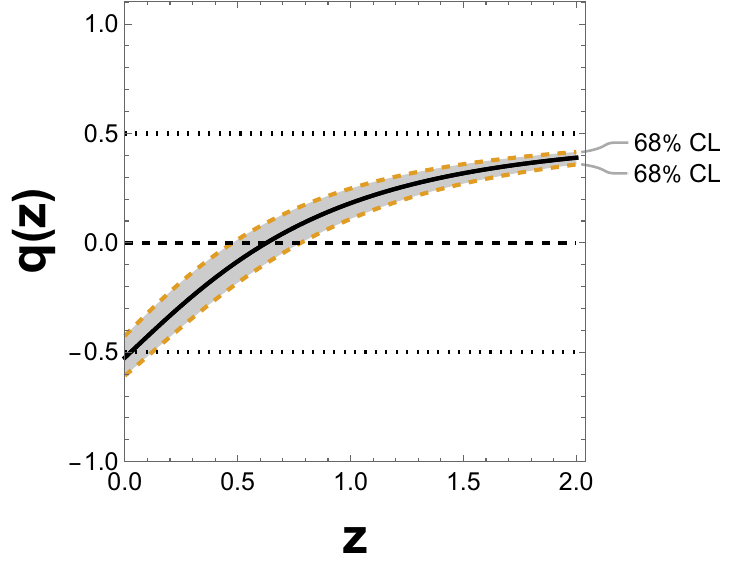} \label{fig:qz68LCDM}}
\caption{The deceleration parameter $q\left(z\right)$ as a function of the redshift. The graph in (\ref{fig:qz68}) represents the evolution of the deceleration parameter $q(z)$ of our model. The continuous line represents the evolution $q(z)$ with the best fit to $\Omega_{m0}=0.297$ and $\beta=0.989$. Dashed lines represent a $68.2\%$ confidence level. The graph in (\ref{fig:qz68LCDM}) represents the evolution deceleration parameter $q(z)$ of $\Lambda\mathrm{CDM}$ model. The continuous line represents the evolution $q(z)$ with the best fit to $\Omega_{m0}=0.317$. Dashed lines represent $68.2\%$ confidence level.}\label{figqztotal}
   \end{figure}

After statistical analysis, the evolution of the deceleration parameter, expression (\ref{eqB:3}), can be seen in Fig.(\ref{fig:qz68}), together with the evolution represented by $68.2\%$ confidence level. We clearly see that the model allows for an accelerated expansion phase with a best fit value for the deceleration parameter today $q_{0}=-0.554$.

\subsection{Model selection criteria}\label{subsec4}

In this paper we use two selection model criteria commonly used in the literature \cite{Naik:2023yhl,delCampo:2011jp} as a  selection method of the best model: the Akaike information criterion (AIC) and the Bayesian information criterion (BIC). 
The AIC method to select the best model is based on the maximum likelihood of the data and the number of free parameters of the model. The expression of the AIC method is

\begin{align}
    AIC=-2ln \mathcal{L}_{max}+2M,
\end{align}
where $\mathcal{L}_{max}$ expresses maximum likelihood; the term $-2ln \mathcal{L}_{max}=\chi^{2}_{min}$ and $M$ is the number of free parameters of the model. The model with the lowest AIC is the one chosen as the reference model, or best model $AIC_{best}$. It is defined after calculating these values for each proposed model. We denote the difference between the result $AIC_{i}$, of some proposed model, with the best model $AIC_{best}$ by $\Delta_{AIC}$. This term is used to determine the support level for the $ith$ model. Therefore, for those models whose $0\leq \Delta_{AIC}\leq 2$, then it means that the $ith$ model is as good as the best model. On the other hand, for those who have $4\leq \Delta_{AIC}\leq 7$, the $ith$ model has considerably less support. While, for $\Delta_{AIC}\geq 10$ indicates the $ith$ model is unlikely and should be discarded.

The second model selection criterion is the BIC. Just like AIC, the BIC criterion penalizes models with more addition of free parameters and favors models that better fit the data. On the other hand, the BIC criterion has a higher penalty than the AIC criterion. The expression of the BIC method is

\begin{align}
    BIC=-2ln \mathcal{L}_{max}+M\,ln N,\label{eq:BIC}
\end{align}
where $N$ is the number of data points used in analysis. Just like on the AIC criterion, the lowest value obtained in expression (\ref{eq:BIC}) is considered the reference model or, the best model ($BIC_{best}$).

\begin{table}[!t] 
 \centering
 \caption{Table containing the results obtained from the final statistical analysis for the cases: $\Lambda \mathrm{CDM}$, $\omega\mathrm{CDM}$ and $\mathrm{HODENUG}$.}
\label{tab:LCDMHDENUG}
\resizebox{15cm}{!}{
\begin{tabular}{|c|c|c|c|}
\hline 
 $Model$ & $\Lambda\mathrm{CDM}$ & $\omega\mathrm{CDM}$ & $\mathrm{HODENUG}$\tabularnewline
\hline 
\hline 
$Best\,\,fit$ & $\Omega_{m0}=0.317;\,\,\,\,h=0.677$ & $\Omega_{m0}=0.332;\,\,\,\,h=0.714;\,\,\,\,\omega=-1.302$ & $\Omega_{m0}=0.297;\,\,\,\,h=0.685;\,\,\,\,\beta=0.989$\tabularnewline
\hline 
$M$ & $2$ & $3$ & $3$\tabularnewline
\hline 
$\chi_{min}^{2}$ & $14.332$ & $14.114$ & $14.306$\tabularnewline
\hline 
$AIC$ & $18.332$ & $20.114$ & $20.306$\tabularnewline
\hline 
$BIC$ & $21.199$ & $24.415$ & $24.607$\tabularnewline
\hline 
$\Delta_{AIC}$ & $0$ & $1.782$ &  $1.974$\tabularnewline
\hline 
$\Delta_{BIC}$ & $0$ & $3.216$ & $3.408$\tabularnewline
\hline 
\end{tabular}
}
\end{table}

We interpret the value of $\Delta_{BIC}$ for an $ith$ model, as evidence that the $ith$ model is the best. Therefore, for those models whose, $0\leq \Delta_{BIC} \leq 2$, this means that it has very weak evidence against the $ith$ model over the best one. On the other hand, for values $2< \Delta_{BIC} \leq 6$ the evidence against $ith$ model is positive. While, for $6<\Delta_{BIC}\leq 10$ the evidence against $ith$ model is strong. Finally, for $\Delta_{BIC}>10$ the evidence against $ith$ model is very strong, this indicates that the ith model is unlikely to be the best one.

To select and evaluate the best model, we calculate the AIC and BIC selection criteria for the $\Lambda\mathrm{CDM}$ model and $\mathrm{HODENUG}$ model. Results can be seen in the table (\ref{tab:LCDMHDENUG}). We can see that in both cases the model with the best fit, that is, the one with the lowest AIC and BIC, is the $\Lambda\mathrm{CDM}$ model. According to the AIC selection criteria in our analysis $\Delta_{AIC}<2$, which indicates that our model is as good as the $\Lambda\mathrm{CDM}$ model. However, according to the BIC selection criterion in our analysis $\Delta_{BIC}=3.408$ this indicates that the evidence against $\mathrm{HODENUG}$ model is positive. We must emphasize that the AIC and BIC model selection criteria penalize those models that have a greater number of free parameters. On the other hand, the $\mathrm{HODENUG}$ and $\omega\mathrm{CDM}$ models have the same number of free parameters, in a direct inspection of the table (\ref{tab:LCDMHDENUG}), we can compare the $\mathrm{HODENUG}$ models and the $\omega\mathrm{CDM}$ model using the same model selection criteria. We see that for this case, the $\omega\mathrm{CDM}$ model has lower AIC and BIC, therefore, it is the reference model. Applying the selection criteria we obtain $\Delta_{AIC}=0.192$ and $\Delta_{BIC}=0.192$, which leads us to conclude that the $\mathrm{HODENUG}$ model is as competitive as the $\omega\mathrm{CDM}$ model.

\newpage
\section{Conclusions}\label{sec6}

By considering the Ricci scalar holographic dark energy density in NUG we recover a model which is similar to a time-dependent cosmological constant $\Lambda(t)$ model. As a result, we arrive at an interaction between holographic dark energy and matter, represented by equations (\ref{eq1:11}) and (\ref{eq1:12}). Cosmological models with interaction are often motivated to solve the cosmic coincidence problem \cite{Zimdahl:2007zz, Velten:2014nra}. On the other hand, models in which the holographic dark energy density behaves like the Ricci scalar also motivated by the same problem \cite{PhysRevD.79.043511}. 

The absence of information to solve the complete set of equations is an intrinsic characteristic of the background cosmological solution in NUG and, consequently, to solve the complete set of equations we need to introduce new information to the system, as detailed in \cite{Fabris_2022, Garcia-Aspeitia:2019yod}. Likewise, when considering a holographic dark energy component in NUG we are still faced with this lack of information. We solve this issue by assuming an interaction term $Q=\beta H \rho_{m}$ in the construction of a background cosmological model, which resulted in a theoretical cosmological model that is viable and leads to a scenario of an accelerated universe expanding described by the Hubble parameter, as given by expression (\ref{eqB:1}). 

Concerning the statistical analysis, we used the CC observational data for $H(z)$, which can be found in the table (\ref{tab:CC}). It is worth mentioning the proposed model has only one extra parameter than the flat $\Lambda\mathrm{CDM}$ model. This features makes this approach very competitive. When we analyze the selection criteria $AIC$ and $BIC$ of the best fit model parameters, provided in the table (\ref{tab:LCDMHDENUG}), the $\Lambda\mathrm{CDM}$ model still prevails as the best model. However, when compared to the $\omega \mathrm{CDM}$ model, which has the same number of free parameters as the proposed model, we reach the following conclusion: according to the selection criteria used, HODENUG is as competitive as the $\omega \mathrm{CDM}$ model.

\bigskip
\noindent
{\bf Acknowledgments:} We thank CNPq, FAPES, FAPEMIG and CAPES for partial financial support.
 
 \bibliography{bibli1}{}

\begin{thebibliography}{10}

\bibitem{Einstein:1916vd}
A.~Einstein, ``{The Foundation of the General Theory of Relativity},'' {\em
  Annalen Phys.}, vol.~49, no.~7, pp.~769--822, 1916.

\bibitem{Dyson:1920cwa}
F.~W. Dyson, A.~S. Eddington, and C.~Davidson, ``{A Determination of the
  Deflection of Light by the Sun's Gravitational Field, from Observations Made
  at the Total Eclipse of May 29, 1919},'' {\em Phil. Trans. Roy. Soc. Lond.
  A}, vol.~220, pp.~291--333, 1920.

\bibitem{LIGOScientific:2016aoc}
B.~P. Abbott {\em et~al.}, ``{Observation of Gravitational Waves from a Binary
  Black Hole Merger},'' {\em Phys. Rev. Lett.}, vol.~116, no.~6, p.~061102,
  2016.

\bibitem{Peebles:2002gy}
P.~J.~E. Peebles and B.~Ratra, ``{The Cosmological Constant and Dark Energy},''
  {\em Rev. Mod. Phys.}, vol.~75, pp.~559--606, 2003.

\bibitem{Zeldovich:1968ehl}
Y.~B. Zel'dovich, A.~Krasinski, and Y.~B. Zeldovich, ``{The Cosmological
  constant and the theory of elementary particles},'' {\em Sov. Phys. Usp.},
  vol.~11, pp.~381--393, 1968.

\bibitem{Copeland:2006wr}
E.~J. Copeland, M.~Sami, and S.~Tsujikawa, ``{Dynamics of dark energy},'' {\em
  Int. J. Mod. Phys. D}, vol.~15, pp.~1753--1936, 2006.

\bibitem{Bull:2015stt}
P.~Bull {\em et~al.}, ``{Beyond $\Lambda$CDM: Problems, solutions, and the road
  ahead},'' {\em Phys. Dark Univ.}, vol.~12, pp.~56--99, 2016.

\bibitem{Martin:2012bt}
J.~Martin, ``{Everything You Always Wanted To Know About The Cosmological
  Constant Problem (But Were Afraid To Ask)},'' {\em Comptes Rendus Physique},
  vol.~13, pp.~566--665, 2012.

\bibitem{RevModPhys.61.1}
S.~Weinberg, ``The cosmological constant problem,'' {\em Rev. Mod. Phys.},
  vol.~61, pp.~1--23, Jan 1989.

\bibitem{Einstein:1919gv}
A.~Einstein, ``{Spielen Gravitationsfelder im Aufbau der materiellen
  Elementarteilchen eine wesentliche Rolle?},'' {\em Sitzungsber. Preuss. Akad.
  Wiss. Berlin (Math. Phys. )}, vol.~1919, pp.~349--356, 1919.

\bibitem{Anderson:1971pn}
J.~L. Anderson and D.~Finkelstein, ``{Cosmological constant and fundamental
  length},'' {\em Am. J. Phys.}, vol.~39, pp.~901--904, 1971.

\bibitem{Ellis:2010uc}
G.~F.~R. Ellis, H.~van Elst, J.~Murugan, and J.-P. Uzan, ``{On the Trace-Free
  Einstein Equations as a Viable Alternative to General Relativity},'' {\em
  Class. Quant. Grav.}, vol.~28, p.~225007, 2011.

\bibitem{Garcia-Aspeitia:2019yod}
M.~A. GarcÍa-Aspeitia, A.~Hernandez-Almada, J.~Magãna, and V.~Motta, ``{The
  Universe acceleration from the Unimodular gravity view point: Background and
  linear perturbations},'' {\em Phys. Dark Univ.}, vol.~32, p.~100840, 2021.

\bibitem{Fabris_2022}
J.~C. Fabris, M.~H. Alvarenga, M.~Daouda, and H.~E.~S. Velten,
  ``Nonconservative unimodular gravity: a viable cosmological scenario?,'' {\em
  The European Physical Journal C}, vol.~82, Jun 2022.

\bibitem{Fabris:2023vop}
J.~C. Fabris, M.~H. Alvarenga, and H.~Velten, ``{Using Cosmological
  Perturbation Theory to Distinguish between GR and Unimodular Gravity},'' {\em
  Symmetry}, vol.~15, no.~7, p.~1392, 2023.

\bibitem{Cohen:1998zx}
A.~G. Cohen, D.~B. Kaplan, and A.~E. Nelson, ``{Effective field theory, black
  holes, and the cosmological constant},'' {\em Phys. Rev. Lett.}, vol.~82,
  pp.~4971--4974, 1999.

\bibitem{Zimdahl:2007zz}
W.~Zimdahl and D.~Pavon, ``{Interacting holographic dark energy},'' {\em Class.
  Quant. Grav.}, vol.~24, pp.~5461--5478, 2007.

\bibitem{Li:2004rb}
M.~Li, ``{A Model of holographic dark energy},'' {\em Phys. Lett. B}, vol.~603,
  p.~1, 2004.

\bibitem{delCampo:2011jp}
S.~del Campo, J.~C. Fabris, R.~Herrera, and W.~Zimdahl, ``{On holographic
  dark-energy models},'' {\em Phys. Rev. D}, vol.~83, p.~123006, 2011.

\bibitem{wang2017holographic}
S.~Wang, Y.~Wang, and M.~Li, ``Holographic dark energy,'' {\em Physics
  reports}, vol.~696, pp.~1--57, 2017.

\bibitem{PhysRevD.79.043511}
C.~Gao, F.~Wu, X.~Chen, and Y.-G. Shen, ``Holographic dark energy model from
  ricci scalar curvature,'' {\em Phys. Rev. D}, vol.~79, p.~043511, Feb 2009.

\bibitem{Fu:2011ab}
T.-F. Fu, J.-F. Zhang, J.-Q. Chen, and X.~Zhang, ``{Holographic Ricci dark
  energy: Interacting model and cosmological constraints},'' {\em Eur. Phys. J.
  C}, vol.~72, p.~1932, 2012.

\bibitem{delCampo:2013hka}
S.~del Campo, J.~C. Fabris, R.~Herrera, and W.~Zimdahl, ``{Cosmology with Ricci
  dark energy},'' {\em Phys. Rev. D}, vol.~87, no.~12, p.~123002, 2013.

\bibitem{Benetti_2019}
M.~Benetti, W.~Miranda, H.~Borges, C.~Pigozzo, S.~Carneiro, and J.~Alcaniz,
  ``Looking for interactions in the cosmological dark sector,'' {\em Journal of
  Cosmology and Astroparticle Physics}, vol.~2019, p.~023, dec 2019.

\bibitem{Zimdahl:2012mj}
W.~Zimdahl, ``{Models of Interacting Dark Energy},'' {\em AIP Conf. Proc.},
  vol.~1471, pp.~51--56, 2012.

\bibitem{1991JMP....32.1337N}
Y.~J. {Ng} and H.~{van Dam}, ``{Unimodular theory of gravity and the
  cosmological constant.},'' {\em Journal of Mathematical Physics}, vol.~32,
  pp.~1337--1340, May 1991.

\bibitem{1993JMP....34.2465T}
S.~C. {Tiwari}, ``{A note on the unimodular theory of gravitation.},'' {\em
  Journal of Mathematical Physics}, vol.~34, pp.~2465--2467, June 1993.

\bibitem{Ellis:2013uxa}
G.~F.~R. Ellis, ``{The Trace-Free Einstein Equations and inflation},'' {\em
  Gen. Rel. Grav.}, vol.~46, p.~1619, 2014.

\bibitem{universe7020038}
H.~Velten and T.~R.~P. Caramês, ``To conserve, or not to conserve: A review of
  nonconservative theories of gravity,'' {\em Universe}, vol.~7, no.~2, 2021.

\bibitem{Susskind:1994vu}
L.~Susskind, ``{The World as a hologram},'' {\em J. Math. Phys.}, vol.~36,
  pp.~6377--6396, 1995.

\bibitem{tHooft:1993dmi}
G.~'t~Hooft, ``{Dimensional reduction in quantum gravity},'' {\em Conf. Proc.
  C}, vol.~930308, pp.~284--296, 1993.

\bibitem{doi:10.1142/S0218271815300074}
Y.~L. Bolotin, A.~Kostenko, O.~A. Lemets, and D.~A. Yerokhin, ``Cosmological
  evolution with interaction between dark energy and dark matter,'' {\em
  International Journal of Modern Physics D}, vol.~24, no.~03, p.~1530007,
  2015.

\bibitem{PhysRevD.73.123510}
B.~Hu and Y.~Ling, ``Interacting dark energy, holographic principle, and
  coincidence problem,'' {\em Phys. Rev. D}, vol.~73, p.~123510, Jun 2006.

\bibitem{Jimenez:2001gg}
R.~Jimenez and A.~Loeb, ``{Constraining cosmological parameters based on
  relative galaxy ages},'' {\em Astrophys. J.}, vol.~573, pp.~37--42, 2002.

\bibitem{29}
C.~Zhang, H.~Zhang, S.~Yuan, S.~Liu, T.-J. Zhang, and Y.-C. Sun, ``Four new
  observational h (z) data from luminous red galaxies in the sloan digital sky
  survey data release seven,'' {\em Research in Astronomy and Astrophysics},
  vol.~14, no.~10, p.~1221, 2014.

\bibitem{31}
M.~Moresco, L.~Pozzetti, A.~Cimatti, R.~Jimenez, C.~Maraston, L.~Verde,
  D.~Thomas, A.~Citro, R.~Tojeiro, and D.~Wilkinson, ``{A 6\% measurement of
  the Hubble parameter at $z\sim0.45$: direct evidence of the epoch of cosmic
  re-acceleration},'' {\em JCAP}, vol.~05, p.~014, 2016.

\bibitem{28}
M.~Moresco, A.~Cimatti, R.~Jimenez, L.~Pozzetti, G.~Zamorani, M.~Bolzonella,
  J.~Dunlop, F.~Lamareille, M.~Mignoli, H.~Pearce, {\em et~al.}, ``Improved
  constraints on the expansion rate of the universe up to z~ 1.1 from the
  spectroscopic evolution of cosmic chronometers,'' {\em Journal of Cosmology
  and Astroparticle Physics}, vol.~2012, no.~08, pp.~006--006, 2012.

\bibitem{27}
D.~Stern, R.~Jimenez, L.~{Verde}, M.~{Kamionkowski}, and S.~A. {Stanford},
  ``{Cosmic chronometers: constraining the equation of state of dark energy. I:
  H(z) measurements},'' {\em jcap}, vol.~2010, p.~008, Feb. 2010.

\bibitem{30}
M.~Moresco, ``Raising the bar: new constraints on the hubble parameter with
  cosmic chronometers at z~ 2,'' {\em Monthly Notices of the Royal Astronomical
  Society: Letters}, vol.~450, no.~1, pp.~L16--L20, 2015.

\bibitem{32}
A.~L. Ratsimbazafy, S.~I. Loubser, S.~M. Crawford, C.~M. Cress, B.~A. Bassett,
  R.~C. Nichol, and P.~V\"ais\"anen, ``{Age-dating Luminous Red Galaxies
  observed with the Southern African Large Telescope},'' {\em Mon. Not. Roy.
  Astron. Soc.}, vol.~467, no.~3, pp.~3239--3254, 2017.

\bibitem{Naik:2023yhl}
D.~M. Naik, N.~S. Kavya, L.~Sudharani, and V.~Venkatesha, ``{Model-independent
  cosmological insights from three newly reconstructed deceleration parameters
  with observational data},'' {\em Phys. Lett. B}, vol.~844, p.~138117, 2023.

\bibitem{Velten:2014nra}
H.~E.~S. Velten, R.~F. vom Marttens, and W.~Zimdahl, ``{Aspects of the
  cosmological \textquotedblleft{}coincidence problem\textquotedblright{}},''
  {\em Eur. Phys. J. C}, vol.~74, no.~11, p.~3160, 2014.

\end{thebibliography}

\end{document}